\documentclass[aps, prd, 10pt,twocolumn,superscriptaddress,nofootinbib]{revtex4-2}
\usepackage[T1]{fontenc}
\usepackage{mathrsfs, amssymb, amsmath, mathtools}  
\usepackage{cancel, comment}
\usepackage{footmisc}
\usepackage{latexsym}
\usepackage{url}
\usepackage{dcolumn}
\usepackage{multirow}
\usepackage{color}
\usepackage{soul}
\usepackage[normalem]{ulem}
\usepackage{amsfonts,amssymb,amsmath, txfonts}
\usepackage{graphicx,epsfig}
\usepackage{psfrag}
\usepackage{hyperref}
\usepackage{mathtools}
\usepackage{enumitem}
\usepackage[dvipsnames]{xcolor}
\usepackage{xcolor}
\hypersetup{linktoc=all,
    colorlinks=true, linkcolor={red},
    citecolor={blue}, urlcolor={purple}
}

\usepackage[percent]{overpic}
\usepackage{slashed}
\usepackage{wrapfig}
\usepackage{tabu}
\usepackage{diagbox}
\usepackage{mathrsfs,amsmath,amssymb,amsfonts,tikz,graphicx,accents,hyperref, color}
\usepackage{dsfont,epiolmec, latexsym, stmaryrd, comment}
\usepackage{slashed,ccaption}
\usepackage{mathrsfs, calligra}
\usepackage{leftidx}
\usepackage{import}
\usepackage{multirow}
\usepackage{amsfonts}
\usepackage{pifont}
\usepackage{tabularx}
\usepackage{cancel}
\usepackage[utf8]{inputenc}
\usetikzlibrary{intersections,calc}
\definecolor{rosy}{RGB}{230,235,252}
\definecolor{myframetitle}{RGB}{90,89,170}
\definecolor{myblocktitle}{RGB}{140,185,249}
\definecolor{mytitle}{RGB}{10,80,26}

\definecolor{darkgreen}{RGB}{27,130,45}
\definecolor{darkblue}{rgb}{0,0,0.3}
\definecolor{darkred}{rgb}{0.7,0,0}

\definecolor{light gray}{RGB}{220,220,220}
\definecolor{dark purple}{RGB}{108,0,217}
\definecolor{pink}{RGB}{190,20,100}
\definecolor{orang}{RGB}{193,63,0}
\definecolor{green}{RGB}{11,98,17}
\definecolor{darkpink}{RGB}{153,0,76}
\definecolor{bluegreen}{RGB}{0,102,102}
\definecolor{greenlagan}{RGB}{0,102,0}
\definecolor{redgreen}{RGB}{102,102,0}
\definecolor{Redgreen}{RGB}{153,76,0}
\definecolor{vividviolet}{rgb}{0.62, 0.0, 1.0}
\definecolor{amaranth}{rgb}{0.9, 0.17, 0.31}
\definecolor{palatinateblue}{rgb}{0.15, 0.23, 0.89}
\definecolor{brightpink}{rgb}{1.0, 0.0, 0.5}
\definecolor{cornflowerblue}{rgb}{0.39, 0.58, 0.93}
\definecolor{deepcarminepink}{rgb}{0.94, 0.19, 0.22}
\definecolor{radicalred}{rgb}{1.0, 0.21, 0.37}
\def\H0{{\text{H}\hspace*{-2.05mm}\text{H} 0\hspace*{-1.35mm}0\ }}

\def\be{\begin{equation}}
\def\ee{\end{equation}}
\def\beq{\begin{equation}}
\def\eeq{\end{equation}}
\def\bea{\begin{eqnarray}}
\def\eea{\end{eqnarray}}
\newcommand{\dd}{\textrm{d}}

\begin{document}

\title{Implications of DES 5YR SNe Dataset for \texorpdfstring{$\Lambda$CDM}{LCDM}}

\author{Eoin \'O Colg\'ain}
\affiliation{Atlantic Technological University, Ash Lane, Sligo F91 YW50, Ireland}
\author{Saeed Pourojaghi}
\affiliation{Department of Physics, Bu-Ali Sina University, Hamedan 65178, Iran}
\author{M. M. Sheikh-Jabbari}
\affiliation{ School of Physics, Institute for Research in Fundamental
Sciences (IPM), P.O.Box 19395-5531, Tehran, Iran}

\begin{abstract}
Dark Energy Survey five-year supernovae data (DES 5YR SNe)  in conjunction with  Planck CMB and Dark Energy Spectroscopic Instrument (DESI) BAO data has detected a strong dynamical dark energy (DE) deviation from the $\Lambda$CDM model.  
Here we shift the focus of DES data to the pressureless matter sector in the $\Lambda$CDM model by studying   
the matter density parameter $\Omega_m$. Employing primarily frequentist profile likelihoods, supported by complementary Bayesian methods, we demonstrate that  $\Omega_m$ increases with effective redshift in the DES data up to a point that there is a $2.5 \sigma$ discrepancy with Planck. We relax the traditional $\Omega_m \leq 1$ prior to demonstrate negative DE densities $\Omega_m > 1$ at the highest effective redshift probed. Nevertheless, the largest discrepancy with Planck occurs for profile likelihoods and posteriors peaked at $\Omega_m < 1$ in the traditional $\Lambda$CDM regime. Our findings corroborate earlier observations in Pantheon and Pantheon+ datasets with an  independent SNe dataset with a higher effective redshift. In an appendix, we confirm that curvature $\Omega_k$ decreases with effective redshift disfavouring a flat Universe in higher redshift DES SNe at $> 3 \sigma$. Our choice of $\Omega_k$ prior leads to an underestimation of the tension with a flat Universe. 

\end{abstract}

\maketitle

\section{Introduction}
In recent years, cosmological or $\Lambda$CDM tensions have caused considerable confusion (see \cite{DiValentino:2021izs, Perivolaropoulos:2021jda, Abdalla:2022yfr} for reviews). Given a physical discrepancy - one not due to systematics - in the Hubble constant $H_0$ assuming the $\Lambda$CDM model, the simplest deduction one can make is that the $\Lambda$CDM  fitting parameter $H_0$ \footnote{$H_0$ is the only cosmological parameter that can be determined model independently just upon the assumption of cosmic homogeneity and isotropy, by extrapolating low redshift data $z \lesssim 0.1$ to $z = 0$. At higher redshifts, one may assume a Taylor expansion in $z$ and the first term is then acceleration parameter. Asking for Hubble parameter at arbitrary redshift $H(z)$ requires a cosmological model.} must not be a constant, i. e. it depends on the redshift at which it is measured \cite{Krishnan:2020vaf, Krishnan:2022fzz}. This begets a research programme where one studies $H_0$ tomographically by confronting the $\Lambda$CDM model to data binned by effective redshift. To that end, descending trends of $H_0$ with redshift in the late Universe have been synergistically reported in the literature for various different datasets \cite{Wong:2019kwg, Millon:2019slk, Krishnan:2020obg, Dainotti:2021pqg, Dainotti:2022bzg, Hu:2022kes, Colgain:2022nlb, Colgain:2022rxy, Jia:2022ycc, jia2024uncorrelated}. 

$H_0$ does not exist in a vacuum; cosmologically, it must talk to other $\Lambda$CDM parameters. The next most relevant parameter is matter density today $\Omega_m$. This parameter is arguably no less special than $H_0$. In the baseline flat $\Lambda$CDM model, $\Omega_m$ parameterises both the pressureless matter and dark energy (DE) sectors. Moreover, it is also the backbone of the $\Lambda$CDM model, describing the evolution of the Universe at the level of equations of motion over 13 billion years through a single parameter. Any textbook on cosmology confirms that the Universe is approximately $30 \%$ matter density corresponding to $\Omega_m \approx 0.3$. If this picture is correct, $\Omega_m$ cannot drift far from $\Omega_m \approx 0.3$. Lastly, $\Omega_m$ is anti-correlated with $H_0$ when one fits data, so the flip side of the decreasing $H_0$ with redshift narrative is an increasing $\Omega_m$ with redshift narrative. Unsurprisingly, claims of increasing values of $\Omega_m$ with effective redshift also exist in the literature \cite{Risaliti:2018reu, Lusso:2020pdb, Colgain:2022nlb, Colgain:2022rxy, Pourojaghi:2022zrh, Pasten:2023rpc}. See \cite{Akarsu:2024qiq} for an overview of the claims.    

An increasing $\Omega_m$ with redshift, if physical, leads one to an ``$\Omega_m$ tension''. The most striking realisation comes from Risaliti-Lusso quasar (QSO) datasets \cite{Risaliti:2015zla, Risaliti:2018reu, Lusso:2020pdb}, which has recently been placed at $\sim 8 \sigma$ level with frequentist methods \cite{colgain2024high} (see also \cite{Yang:2019vgk, Khadka:2020vlh, Khadka:2020tlm}). Coupled with $H_0$ tension, a putative $\sim 5 \sigma$ tension \cite{Riess:2021jrx}, one could be looking at an existential crisis for the vanilla $\Lambda$CDM model. Understandably, given the strong disagreement with the standard model, the QSO dataset has been heavily criticised \cite{Khadka:2020vlh, Khadka:2020tlm, Khadka:2021xcc, Khadka:2022aeg, Singal:2022nto, Petrosian:2022tlp, Zajacek:2023qjm}. Nevertheless, the observation that $\Omega_m$ increases with effective redshift in the flat $\Lambda$CDM model extends to Type Ia SNe \cite{Colgain:2022nlb, Colgain:2022rxy, Malekjani:2023dky}, in particular Pantheon \cite{Pan-STARRS1:2017jku} and Pantheon+ \cite{Scolnic:2021amr, Brout:2022vxf} samples. The problem with these samples is that at higher redshifts the samples become sparse and the $z > 1$ SNe are common to Pantheon and Pantheon+ \footnote{Pantheon+ cuts 57 high redshift SNe from the SNLS sample \cite{SNLS:2011lii, SNLS:2011cra} due to potential evolution in inferred distances \cite{Brout:2021mpj}, so Pantheon+ statistics are worse than Pantheon at higher redshifts.}, preventing one from expanding the science case. 

The Dark Energy Survey's five year (DES 5YR) SNe sample \cite{DES:2024tys, DES:2024upw,the_des_sn_working_group_2024_12720778} offers considerable promise. First, in contrast to QSOs, Type Ia SNe are better quality \textit{standardisable} candles with greater community trust. Secondly, the DES sample has better high redshift statistics compared to Pantheon+ \cite{Scolnic:2021amr, Brout:2022vxf} or Union3 \cite{Rubin:2023ovl}. Thirdly, deviations from $\Lambda$CDM behaviour have already been reported by the DES collaboration in the DE sector \cite{DES:2024tys}, a problem that is exacerbated by combining the dataset with DESI BAO \cite{DESI:2024mwx} and Planck CMB constraints \cite{Planck:2018vyg}. It is worth noting that between Pantheon+, Union3 and DES, the CMB+BAO+SNe signal for dynamical DE is $2.5 \sigma, 3.5 \sigma$ and $3.9\sigma$ level, respectively \cite{DESI:2024mwx}. In DES data, this signal is driven by differences in the $\Lambda$CDM $\Omega_m$ parameter with the DE sector compensating, as is evident from Fig. 8 of \cite{DES:2024tys}.\footnote{Puzzlingly, $w(z=0) = w_0$ is consistent with zero within $1 \sigma$ \cite{DES:2024tys}, thus problematic for late-time accelerated expansion today. Moreover, the Universe is younger by $9\%$ or $1.3$ gigayears with dynamical DE \cite{DES:2024tys}.}

In this letter, we confirm existing results in the literature that $\Omega_m$ increases with effective redshift in the \textit{independent} DES 5YR sample. Noting that the parameter $\Omega_m$ is common to both the DE and pressureless matter sectors, neglecting a DES SNe systematic, this implies that the $\Lambda$CDM cosmology is incorrectly modelling one if not both of the sectors. One can of course decide by \textit{fiat} that the missing physics is confined to the DE sector, which is the prevailing assumption in the literature, e. g. DES analysis \cite{DES:2024tys}. However, this deduction may not be supported by our study here, or previous studies \cite{Colgain:2022nlb, Colgain:2022rxy, Malekjani:2023dky}, because the deviations we see are driven by higher redshift data, which is more a probe of matter dominated physics than DE dominated physics.\footnote{There is a common misconception that studying constraints on the $\Lambda$CDM parameter $\Omega_m$ in redshift bins is equivalent to studying a $w(z)$CDM model with varying DE equation of state $w(z)$. This is mathematically incorrect because the $\Lambda$CDM parameter $\Omega_m$ is shared by both the DE and matter sectors, whereas $w(z)$ only models the DE sector by definition.} Ultimately, the surprise is not that $\Omega_m$ varies between CMB, BAO and SNe datasets \cite{DESI:2024mwx}, but that it can vary within a single (SNe) dataset when it is analysed tomographically.

\section{Methodology}\label{sec:methods}
Here, we  use frequentist confidence intervals based on profile likelihood ratios (see \cite{Trotta:2017wnx} for a review) to track differences in cosmological fitting parameters. We  use two methodologies. The first is based on Wilks' theorem \cite{Wilks} and assumes that the profile likelihood is Gaussian, an assumption that is valid in the large sample limit; the DES SNe sample is large and it is expected that one can safely apply Wilks' theorem where the profile likelihoods are demonstrably Gaussian. The second is based on a method that is applicable to non-Gaussian profile likelihoods \cite{Gomez-Valent:2022hkb, Colgain:2023bge, colgain2024high}, yet is guaranteed to recover results consistent with Wilks' theorem in the Gaussian regime. We later confirm our results using Bayesian Markov Chain Monte Carlo (MCMC) methods. 

On the data side, we analyse the DES 5YR sample \cite{DES:2024tys} of 1829 SNe, 194 of which are common to CfA \cite{Hicken:2009df, Hicken:2012zr}, CSP \cite{Krisciunas:2017yoe} and Foundation samples \cite{Foley:2017zdq}, leaving 1635 independent SNe in the DES sample. What interests us here is that the sample has a higher effective redshift than Pantheon+ \cite{Scolnic:2021amr, Brout:2022vxf}. In short, we repeat the analysis of  \cite{Malekjani:2023dky} but with the independent DES sample with better statistics at higher redshifts.    

We are interested in the flat $\Lambda$CDM model with Hubble parameter,  
\begin{equation}
    H(z) = H_0 \sqrt{1-\Omega_m + \Omega_m (1+z)^3},   
\end{equation}
where $H_0$ is the Hubble constant and $\Omega_m$ is the matter density parameter. We treat both as fitting parameters. From here, one constructs the luminosity distance  $D_{L}(z)$
\begin{equation}\label{DLz-DMz}
    D_{L}(z)=\frac{c}{H_0} (1+z) \chi(z),\qquad  \chi(z):=\int_0^{z} \frac{H_0}{H(z^{\prime})} \dd z^{\prime},
\end{equation}
%$ = c (1+z) D_{M}(z)$ from the comoving distance: 
which appears in the distance modulus: 
\begin{equation}
    \mu(z) = 5 \log_{10} D_{L}(z) + 25. 
\end{equation}

To fit DES 5YR data, we consider the $\chi^2$ -likelihood: 
\begin{equation}
    \label{eq:chi2}
    \chi^2 = \Delta \mu_i C_{ij}^{-1} \Delta \mu_j, 
\end{equation}
where $C_{ij}$ is the $1829 \times 1829$ covariance matrix incorporating statistical and systematic uncertainties \cite{DES:2024tys} and $\Delta \mu_i = \mu(z_i)- \mu_i$ is the difference between the model prediction at redshift $z_i$ and the observed distance modulus $\mu_i$. The DES collaboration provide Hubble diagram redshifts, distance moduli and the corresponding distance moduli errors $\sigma_{\mu_i}$ \cite{DES:2024tys, DES:2024upw, the_des_sn_working_group_2024_12720778}. We added $\sigma_{\mu_i}^2$ to the diagonal of the covariance matrix provided by DES. This is validated later by recovering DES results. 

At this juncture, we depart from DES analysis by employing profile likelihoods in place of MCMC. As we shall see, for the full sample, either with or without 194 external SNe \cite{Hicken:2009df, Hicken:2012zr, Krisciunas:2017yoe, Foley:2017zdq}, we recover DES results. This provides an important consistency check on our frequentist methods. Profile likelihoods for one degree of freedom $\theta$ are constructed by isolating one of the parameters $\theta \in \{H_0, \Omega_m \}$ and scanning over it while maximising the likelihood $\mathcal{L} \propto e^{-\frac{1}{2} \chi^2}$ with respect to the other parameter. Here, we focus on $\Omega_m$ and maximise with respect to $H_0$. How $H_0$ is calibrated is not an issue, it is simply an auxiliary (nuisance) parameter in the profile likelihood analysis. The profile likelihood ratio for $\Omega_m$ is \cite{Trotta:2017wnx}
\begin{equation}
    \label{eq:R}
    R(\Omega_m) = \exp \left( -\frac{1}{2} ( \chi^2_{\textrm{min}}(\Omega_m) - \chi^2_{\textrm{min}}) \right),  
\end{equation}
where $\chi^2_{\textrm{min}} (\Omega_m)$ is the minimum of the $\chi^2$, or alternatively maximum of the likelihood, for each value of $\Omega_m$, whereas $\chi^2_{\textrm{min}}$ is the global minimum of the $\chi^2$ over all values of $\Omega_m$ that we probe. The reader will note that the profile likelihood ratio (\ref{eq:R}) peaks at $R(\Omega_m) = 1$ by construction. 

With the profile likelihood ratio constructed, we  employ two different methods to identify confidence intervals. The first \textit{assumes} that the profile likelihood is Gaussian in the large sample limit \cite{Wilks}. One then identifies $100 \,\alpha \%$ confidence intervals corresponding to the values of $\Delta \chi^2$ satisfying \cite{Trotta:2017wnx}: 
\begin{equation}
\label{eq:alpha}
    \alpha = \int_{y=0}^{y = \Delta \chi^2} \frac{1}{\sqrt{2 \pi y}} e^{-\frac{1}{2} y} \textrm{d} y, 
\end{equation}
where we have specialised to the chi-squared distribution with one degree of freedom and employed $\Gamma(\frac{1}{2}) = \sqrt{\pi}$. Integrating the right hand side to $\Delta \chi^2 = 1$ and $\Delta \chi^2 = 3.9$ one finds $\alpha = 0.6827$ and $\alpha = 0.9517$, corresponding to $68 \%$ and $95 \%$ confidence intervals, respectively. 

Alternatively, one can construct, $68\%, 95\%$ confidence intervals for the profile likelihood, by normalising the profile likelihood ratio by the total area under the curve \cite{Gomez-Valent:2022hkb, Colgain:2023bge}, 
\be
\label{eq:w}
w(\Omega_m) = \frac{R(\Omega_m)}{\int R(\Omega_m) \, \dd \Omega_m}, 
\ee
and solving the equation 
\be
\label{eq:conf}
\int_{\Omega_m^{(1)}}^{\Omega_m^{(2)}} w(\Omega_m) \, \dd \Omega_m = \{0.68, 0.95\}, \quad w(\Omega_m^{(1)}) = w(\Omega_m^{(2)}). 
\ee
It should be noted that (\ref{eq:w}) and (\ref{eq:conf}) are essentially equations (2) and (3) from \cite{Herold:2021ksg}. The content of  (\ref{eq:conf}) is that one builds confidence intervals outwards from the peak of the profile likelihood ratio or maximum likelihood estimator (MLE) by steadily including points in parameter space that are incrementally less likely until one reaches $68 \%, 95 \%$, etc, of the area under the profile likelihood curve. It is apparent from the construction that when the profile likelihoods are close to Gaussian the two methods agree. We will show this explicitly for our case, but further explanations can be found in \cite{colgain2024high}.

We remark that there is a Feldman-Cousins prescription \cite{Feldman:1997qc} for handling boundaries, i. e. priors imposed on parameters. Since the MLE or peak of our profile likelihood $R(\Omega_m)$ is located at larger $\Omega_m$ values than the Planck value $\Omega_m \approx 0.3$, and the boundary is at $\Omega_m = 0$, the boundary never impacts our lower confidence intervals or the inferred tensions. Thus, the Feldman-Cousins prescription is irrelevant. However, we do relax the usual $\Omega_m \leq 1$ prior in order to identify the MLE for $z> 0.8$ SNe. Note, while General Relativity allows for a cosmological constant $\Lambda$, there is no theoretical argument that $\Lambda$ needs be positive (recall that in $\Lambda$CDM, $\Lambda\propto (1-\Omega_m)$). In fact, a small positive $\Lambda$ causes well-documented problems for both quantum field theory \cite{Weinberg:1988cp} and string theory \cite{Dvali:2014gua, Dvali:2018fqu, Obied:2018sgi}. For us $\Lambda$CDM is defined with $\Omega_m\geq 0$ bound, allowing for $\Omega_m>1$. Our definition makes a good theoretical sense as there is no strong theoretical reason for $\Omega_\Lambda:=1-\Omega_m$ to be positive, see e.g. \cite{Weinberg:1988cp, Dvali:2014gua, Dvali:2018fqu, Obied:2018sgi}, whereas $\Omega_m\geq 0$ (the weak energy condition) has a theoretical backing. 
Relaxing the traditional $\Omega_m \leq 1$ prior allows one to model a negative $\Lambda$ in the $\Lambda$CDM model.\footnote{In this letter we use the Planck $\Omega_m$ value, $\Omega_m = 0.315 \pm 0.007$ \cite{Planck:2018vyg} as a ``yardstick" simply to quantify shifts in likelihoods. Being a yardstick, one is free to replace it with another $\Omega_m$ constraint and it will not change the outcome. All that matters is that the yardstick is independent of DES SNe. We also note that despite Planck formally imposing an $\Omega_m \leq 1$ prior, which we relax, there is no contradiction from making any comparison because $\Omega_m$ values in the Planck baseline  MCMC chains never venture beyond the range $0.29 \lesssim \Omega_m \lesssim 0.34$. Thus, repeating Planck analysis allowing for $\Omega_m > 1$, where $\Omega_m$ is a derived parameter that is not directly fitted to the data, noting that the CMB data itself places very strong restrictions on $\Omega_m$, one expects to get back the same results; the data is so restrictive that the Planck $\Omega_m$ value is $98 \sigma$ from the prior $\Omega_m \leq 1$. Obviously, one cannot be $98 \sigma$ from the upper bound unless the $\Omega_m$ values in the MCMC chain are far from the upper bound.}

\section{Results}\label{sec:results}

\begin{figure}[htb]
   \centering
\includegraphics[width=90mm]{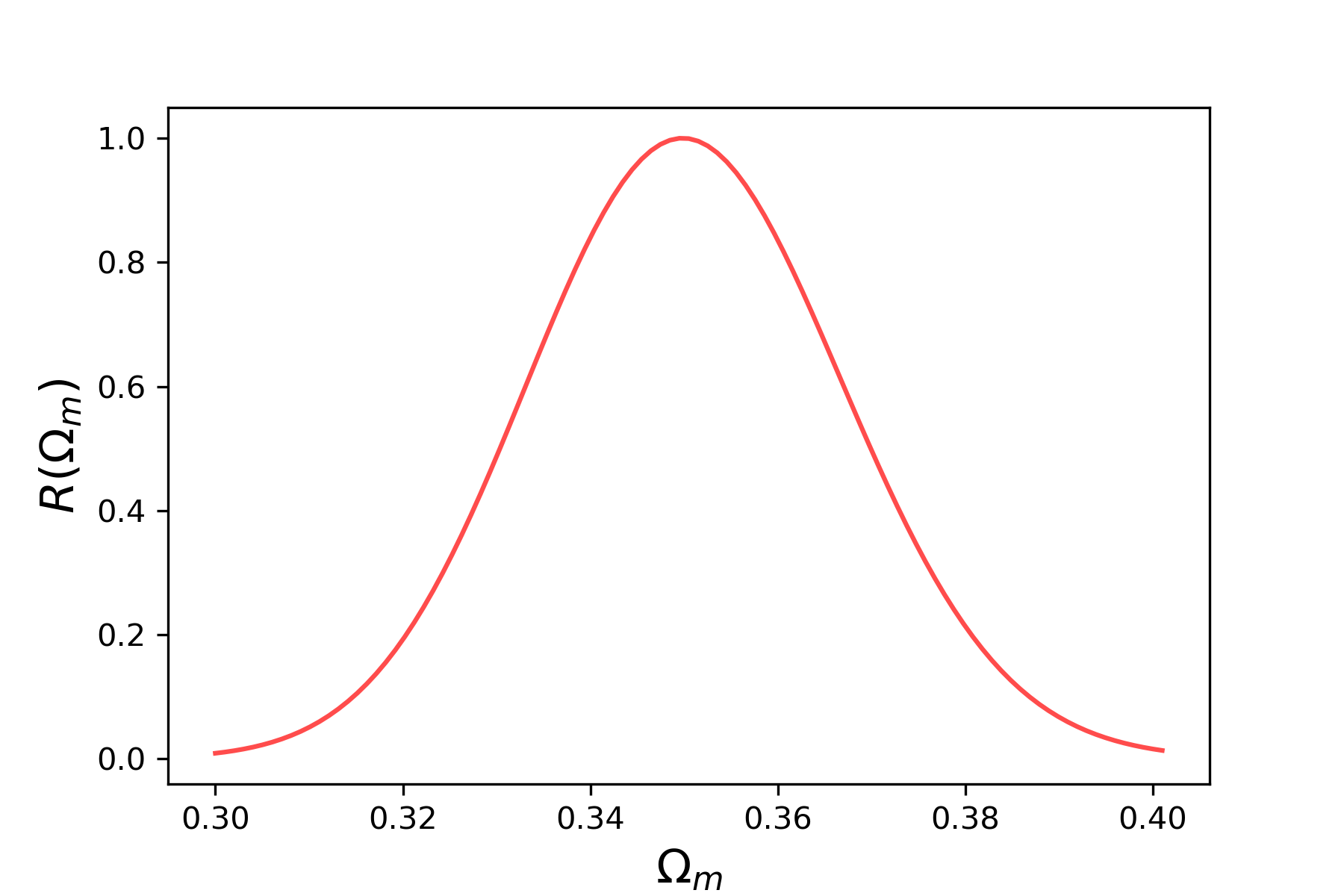} 
\caption{The $\Omega_m$ profile likelihood ratio for the full DES 5YR SNe sample of 1829 SNe.}
\label{fig:DES_BF}
\end{figure}

As a consistency check on our methodology, we first construct the profile likelihood for the full DES sample of 1829 SNe and recover $\Omega_m = 0.352 \pm 0.017$ value from Table 2 \cite{DES:2024tys}. Concretely, whether we employ the approximation $\Delta \chi^2 \leq 1$ or integrate under the profile likelihood curve following  (\ref{eq:w}) and (\ref{eq:conf}), we find $\Omega_m = 0.349^{+0.017}_{-0.016}$. The profile likelihood ratio is illustrated in Fig. \ref{fig:DES_BF}, where we have minimised the $\chi^2$ in the range $\Omega_m \in [0.3, 0.4]$ in steps of $\Delta \Omega_m = 0.01$. Note, we have discretised the $\Omega_m$ parameter space and scanned over it, so our central value is accurate up to the $\Delta \Omega_m = 0.01$ precision. Nevertheless, the important point here is that we agree with DES on the errors with independent methodology, more precisely profile likelihoods in place of MCMC marginalisation. This provides reassurance that we have properly propagated statistical and systematic uncertainties through the DES covariance matrix.

\begin{figure}[htb]
   \centering
\includegraphics[width=90mm]{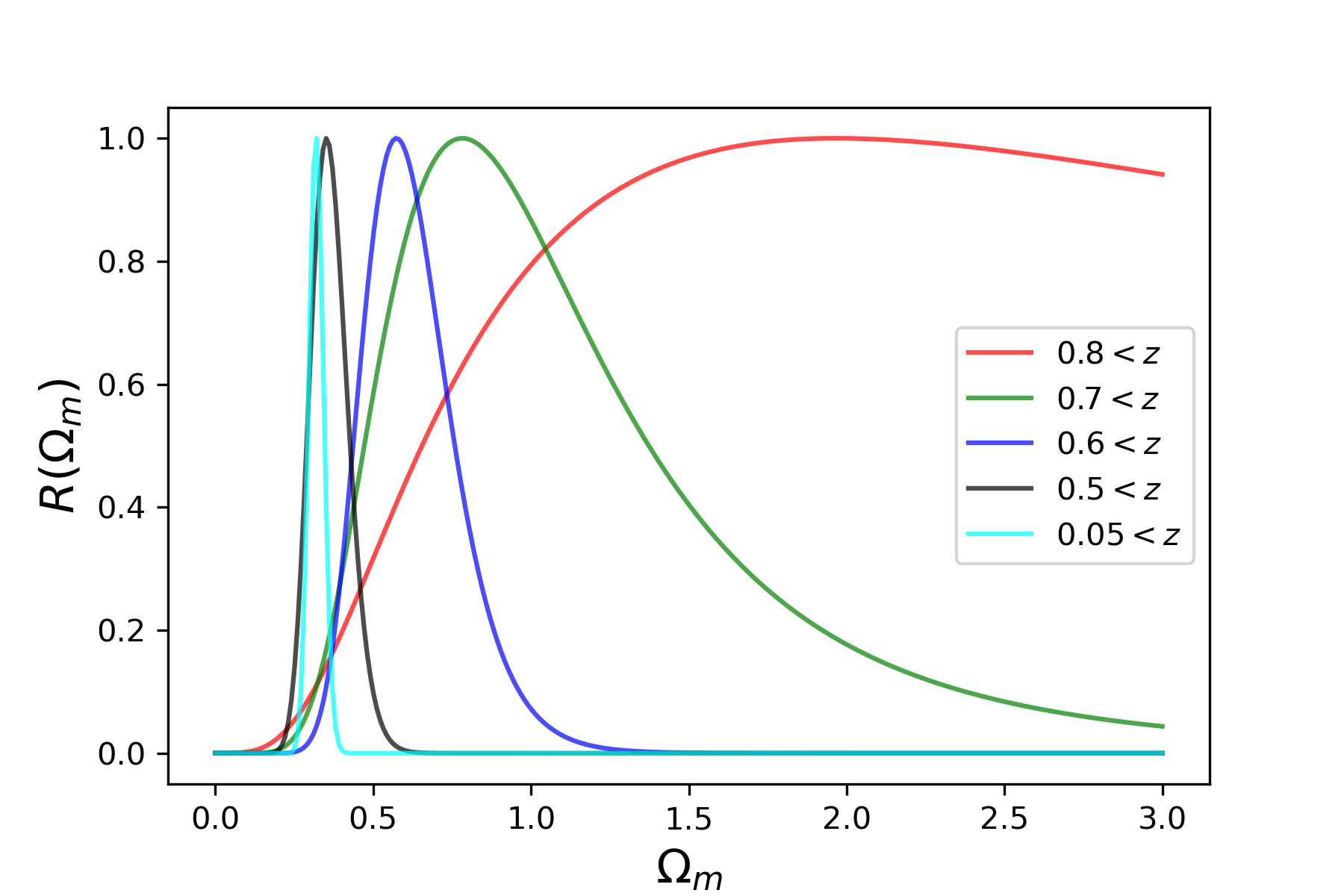} 
\caption{Tomographic or redshift-binned profile likelihood ratios. The MLE shifts to larger values with increasing effective redshift. For the highest redshift range the MLE is in the negative DE density regime with $\Omega_m > 1$. }
\label{fig:DES_OM}
\end{figure}

Our next step follows \cite{Malekjani:2023dky} and involves truncating out the lowest redshift SNe and repeating the analysis. We focus on approximately 300 $\Omega_m$ values evenly spaced in the range $\Omega_m \in [0, 3]$. We start by removing the 194 SNe from datasets \cite{Hicken:2009df, Hicken:2012zr, Krisciunas:2017yoe, Foley:2017zdq} external to DES and crop the covariance matrix accordingly. This leads to the first entry in Table \ref{tab:OMvsZ}. Interestingly, the removal of external data shifts $\Omega_m$ back closer to the Planck value \cite{Planck:2018vyg}. Nevertheless, as is evident from Table \ref{tab:OMvsZ} and Fig. \ref{fig:DES_OM}, removing the lowest redshift SNe from exclusively DES data leads to larger values of $\Omega_m$. In particular, already for $z > 0.5$ SNe we recover the higher $\Omega_m \approx 0.35$ value preferred by the full sample. In addition, for $z > 0.6$ SNe, a discrepancy is evident with canonical Planck values. We emphasise here again that the profile likelihood is visibly Gaussian, a feature expected in the large sample limit, thereby guaranteeing that Wilks' theorem \cite{Wilks} applies.

From Table \ref{tab:OMvsZ} and Fig. \ref{fig:DES_OM}, it is evident that our two methods for constructing frequentist confidence intervals agree well provided the profile likelihoods are close to Gaussian and $R(\Omega_m)$ falls off to small values within our priors. Nevertheless, of the two methods, confidence intervals based on  (\ref{eq:w}) and (\ref{eq:conf}) are more conservative. Evidently, the profile likelihoods become broader and more non-Gaussian at higher redshifts. We note that $\Delta \chi^2 \leq 1$ corresponds to $R(\Omega_m) \geq 0.607$, so one can eyeball the confidence intervals by drawing a horizontal line in Fig. \ref{fig:DES_OM} at $R(\Omega_m) \approx 0.6$. It is clear that the red profile likelihood curve only intersects $R(\Omega_m) \approx 0.6$ once within our bounds, so we only quote lower bounds in Table \ref{tab:OMvsZ}. Anyone surprised by the result should look at Fig. 8 of \cite{DES:2024tys}, where it is evident that a dynamical DE sector is compensating larger $\Omega_m$ values. Our findings here offer greater transparency.

\begin{table}[htb]
\centering 
\begin{tabular}{c|c|c|c}
 \rule{0pt}{3ex} Redshift & \# SNe & $ \Omega_{m}$ ($68 \%$) & $\Omega_m$ ($\Delta \chi^2 \leq 1$) \\
\hline 
\rule{0pt}{3ex} $0.05 < z$ & $1635$ & $0.321^{+0.020}_{-0.020}$ & $0.321^{+0.020}_{-0.020}$ \\
\hline
\rule{0pt}{3ex} $0.5 < z$  & $874$ & $0.351^{+0.070}_{-0.050}$ & $0.351^{+0.060}_{-0.050}$ \\
\hline
%\rule{0pt}{3ex} $0.6 < z$ & $554$ & $0.572^{+0.170}_{-0.120}$ & $0.572^{+0.151}_{-0.110}$ \\
\rule{0pt}{3ex} $0.6 < z$ & $554$ & $0.57^{+0.17}_{-0.12}$ & $0.57^{+0.15}_{-0.11}$ \\
\hline
%\rule{0pt}{3ex} $ 0.7 < z$ & $259$ & $0.783^{+0.592}_{-0.311}$ & $0.783^{+0.472}_{-0.271}$  \\
\rule{0pt}{3ex} $ 0.7 < z$ & $259$ & $0.78^{+0.59}_{-0.31}$ & $0.78^{+0.47}_{-0.27}$  \\
\hline
%\rule{0pt}{3ex} $ 0.8 < z$ & $128$ & $1.385<$ ($1.967$) & $0.763<$ ($1.967$)  \\
\rule{0pt}{3ex} $ 0.8 < z$ & $128$ & $1.39<$ ($1.97$) & $0.76<$ ($1.97$)  \\
\end{tabular}
\caption{$68 \%$ confidence intervals inferred with two methods from DES SNe in different redshift ranges. The number of SNe are also documented. For one-sided confidence intervals we record the MLE in brackets.}
\label{tab:OMvsZ}
\end{table}

We can work out the disagreement with Planck for the $1 \sigma$ upper bound $\Omega_m \leq 0.322$ \cite{Planck:2018vyg}. We do this by extracting $\Delta \chi^2$ values and integrating the profile likelihood. Focusing on $\Delta \chi^2$, we identify the maximum value of $R(\Omega_m)$ corresponding to the Planck $1 \sigma$ upper bound. The 1635 DES SNe sample in cyan in Fig. \ref{fig:DES_OM} is evidently within the Planck $1 \sigma$ confidence interval and only $\Delta \chi^2 = 0.1$ removed from the Planck central value $\Omega_m = 0.315$ \cite{Planck:2018vyg}, which makes it consistent at $0.3 \sigma$. For DES SNe with redshifts $z > 0.8$ (red curve), the Planck $1 \sigma$ upper bound corresponds to $R(\Omega_m) \leq 0.111 \Leftrightarrow \Delta \chi^2 \geq 4.4$, which disfavours the Planck value at $96.4\%$ confidence level or $\sim 2.1 \sigma$. For DES SNe with redshifts $z > 0.7$ (green curve),  the Planck values correspond to $R(\Omega_m) \leq 0.109 \Leftrightarrow \Delta \chi^2 \geq 4.4$, which again disfavours Planck at $\sim 2.1 \sigma$. For DES SNe with redshifts $z > 0.6$ (blue curve), we find $R(\Omega_m) \leq 0.045 \Leftrightarrow \Delta \chi^2 \geq 6.2$, which disfavours Planck at $98.7 \%$ confidence level or $\sim 2.5 \sigma$. Finally, for DES SNe with redshifts $z > 0.5$, we have $R(\Omega_m) \leq 0.858 \Leftrightarrow \Delta \chi^2 \geq 0.3$, which is only $\sim 0.5 \sigma$ removed from Planck. It is worth noting that from the redshift ranges sampled, we find the largest discrepancy with Planck from a profile likelihood that is close to Gaussian, thus any deviation from Wilks' theorem is expected to be small. Finally, to avoid confusion a comment is in order. While the profile likelihoods in Fig. \ref{fig:DES_OM} are correlated because of shared high redshift SNe, Planck CMB data \cite{Planck:2018vyg} is an independent and thus \textit{uncorrelated} dataset, thereby providing a yardstick to quantify the evolution in $\Omega_m$.

As a further aside, although we relax the traditional $\Lambda$CDM bound $\Omega_m \leq 1$ to allow the profile likelihood (later posterior) peaks to enter the $\Omega_m > 1$ regime corresponding to negative dark energy densities, this is only an issue for the red curve in Fig. \ref{fig:DES_OM}. At lower redshifts, the peaks inhabit the traditional range $0 \leq \Omega_m \leq 1$. Thus, neglecting the red curve, any discrepancies with Planck are genuine discrepancies that assume the usual definition of the $\Lambda$CDM model. Note, the largest discrepancy $\sim 2.5 \sigma$ is found for the blue curve in Fig. \ref{fig:DES_OM}.

We repeat the analysis by dropping the assumption of a Gaussian profile likelihood and integrate under the curve following  (\ref{eq:w}) and (\ref{eq:conf}). The full DES SNe sample (cyan curve) we find to be consistent with the Planck central value $\Omega_m = 0.315$ at $16.7 \%$ confidence level or $ \sim 0.2 \sigma$. For DES SNe with $z > 0.8$ (red curve), the Planck $1 \sigma$ upper bound appears at $99.6 \%$ confidence level or $\sim 2.9 \sigma$. We note that this number differs considerably from $\sim 2.1 \sigma$ from Wilks' theorem, but this shows the impact of non-Gaussianities and the boundary. We anticipate that one can relax the $\Omega_m \in [0, 3]$ bound and this would reveal a profile likelihood ratio that falls away gradually for larger $\Omega_m$ values. What this will 
mean is that the area under the profile likelihood curve consistent with Planck $\Omega_m$ values or smaller will be even less of the overall area, thus disfavouring Planck more. For DES SNe with $z > 0.7$ (green curve), the Planck $1 \sigma$ upper bound is disfavoured at $95.4 \%$ confidence level or $\sim 2 \sigma$. This compares favourably with $\sim 2.1 \sigma$ from Wilks' theorem. For DES SNe with $z > 0.6$, the Planck $1 \sigma$ upper bound is disfavoured at $98.4 \%$ confidence level or $\sim 2.4 \sigma$, compared to $\sim 2.5 \sigma$ from our earlier analysis based on Wilks' theorem. We emphasise again that for close to Gaussian profile likelihoods the agreement is good. Finally, for $z > 0.5$ DES SNe, we find the Planck $1 \sigma$ upper bound appears at $38.8 \%$ confidence level or $\sim 0.5 \sigma$. 

\begin{figure}[htb]
   \centering
\includegraphics[width=80mm]{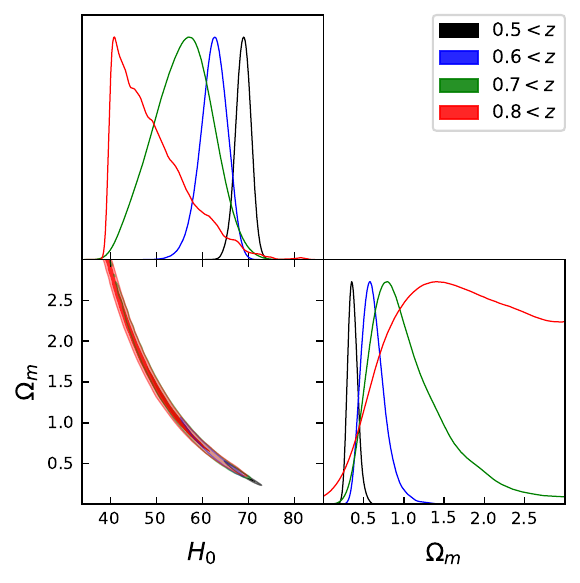} 
\caption{Same as Fig. 2 but profile likelihoods replaced by MCMC posteriors. We have removed the cyan curve as it is poorly differentiated from the black curve.}
\label{fig:DES_corner}
\end{figure}

While frequentist methods may be unfamiliar, in Fig. \ref{fig:DES_corner} we illustrate 2D MCMC posteriors, where the same increasing $\Omega_m$ trend with effective redshift is evident. The reader should note the similarity between Fig. \ref{fig:DES_OM}/Fig. \ref{fig:DES_corner} and Fig. 3 of \cite{colgain2024high}, where SNe are replaced with QSOs with completely different systematics, yet the trend is the same. Given that Gaussian profile likelihoods are protected by Wilks' theorem \cite{Wilks} in the large sample limit; this is the expected outcome. Visually, one notes that the peaks of the profiles and posteriors occur at approximately the same $\Omega_m$ values, except for the red curves as the 2D red posterior stretches to the point that it is impacted by the upper bound $\Omega_m \leq 3$. Once this happens, one no longer expects  generically \textit{prior-independent} frequentist methods and Bayesian methods, which are by construction \textit{prior-dependent},  to agree. The final lesson from Fig. \ref{fig:DES_corner} is that, since $H_0$ is anti-correlated with $\Omega_m$, an increasing $\Omega_m$ trend with effective redshift is expected to imply a decreasing $H_0$ trend with effective redshift. The latter observation can be traced to 2020 \cite{Krishnan:2020obg}.\footnote{See also earlier papers \cite{Wong:2019kwg, Millon:2019slk}. Note that a redshift-dependent $H_0$ from strong lensing time delay undermines the claimed $H_0$ determination.} This is evident in the MCMC posteriors, but not in the profile likelihoods as one extremises with respect to the nuisance parameter $H_0$.\footnote{The DES collaboration makes a specific choice for the $H_0$ calibration. The reader should not focus on the absolute $H_0$ value, but relative differences.}

The main takeaway message from our analysis is that $\Omega_m$ is not a constant in DES data, and as is evident from Fig. \ref{fig:DES_OM} and Table \ref{tab:OMvsZ}, it increases with effective redshift. It should be evident from Fig. 8 in \cite{DES:2024tys} that a larger $\Omega_m$ value drives the dynamical DE signal. It is important to stress again that if DES reports a $\gtrsim 2 \sigma$ deviation from $\Lambda$CDM when it is embedded in the $w_0 w_a$CDM \cite{DES:2024tys} and we found no evolution in $\Omega_m$ in the baseline model when the data is binned by redshift, this would constitute a contradiction. We observe that the $\gtrsim 2 \sigma$ effects we see are of the same statistical significance as the $\Lambda$CDM deviations reported by DES. More generally, one can map any $w_0 w_a$CDM cosmology back to the flat $\Lambda$CDM model parameter $\Omega_m$ as outlined in \cite{Colgain:2024mtg}. For this reason, deviations from $\Lambda$CDM behaviour must be evident in both descriptions, namely in the $w_0 w_a$CDM model and the flat $\Lambda$CDM model confronted to redshift-binned data.

\section{Discussion}
We remind the reader again that persistent $\Lambda$CDM tensions \cite{DiValentino:2021izs, Perivolaropoulos:2021jda, Abdalla:2022yfr}, in particular $H_0$ tension \cite{Riess:2021jrx}, necessitate one finding changes in $\Lambda$CDM fitting parameters with effective redshift \cite{Krishnan:2020vaf, Krishnan:2022fzz}. This is an irrefutable signature of model breakdown. As demonstrated recently in \cite{colgain2024high} with similar frequentist methods, Risaliti-Lusso QSOs and gamma-ray burst (GRB) datasets exist that support $\Omega_m$ increasing tomographically with redshift in the $\Lambda$CDM model. Nevertheless, at best, QSOs and GRBs are emerging cosmological probes \cite{Moresco:2022phi}, and their standardisability is too easy to question. 

This is much more difficult to do with Type Ia SNe, the gold standard of cosmological distance indicators. Decreasing $H_0$/increasing $\Omega_m$ trends have been observed in Type Ia SNe \cite{Krishnan:2020obg, Dainotti:2021pqg, Dainotti:2022bzg, Hu:2022kes, Colgain:2022nlb, Colgain:2022rxy, Jia:2022ycc, jia2024uncorrelated, Pourojaghi:2022zrh, Pasten:2023rpc}, in particular in Pantheon \cite{Pan-STARRS1:2017jku} and Pantheon+ \cite{Scolnic:2021amr, Brout:2022vxf} samples. The problem therein is that at the highest redshifts the statistics are low. The new DES 5YR sample plugs this gap, yet in this letter we find a qualitatively similar increasing $\Omega_m$ with effective redshift trend to QSOs/GRBs. We stress that our observations here should be unsurprising. Fig. 8 of \cite{DES:2024tys} is evidently a funky plot; {it should be obvious that a larger $\Omega_m$ value is compensated by dynamical DE with potentially curious side effects, namely a lack of late-time accelerated expansion in Table 2 \cite{DES:2024tys}, $w(z=0)=w_0 \sim 0$ within $1 \sigma$, and a Universe younger by $9\%$ \cite{DES:2024tys}.} We emphasise again that while we have adopted different methodology, we recovered DES results for the full sample. One could attribute the increasing $\Omega_m$ trend with effective redshift to systematics in the DES sample \footnote{See \cite{Efstathiou:2024xcq} for comments on how a global offset in apparent magnitude $m$ could improve consistency with the Planck-$\Lambda$CDM model. Note, all our analysis here is binned, i. e. local, so a global offset cannot address it. All a global offset in $m$ does is shift the degenerate $H_0$ nuisance parameter in our profile likelihood analysis.} or problems standardising Type Ia SNe at higher redshifts, however, given the difference in astrophysics with QSOs/GRBs, yet the same result \cite{colgain2024high}, this demands that scrutiny falls on the $\Lambda$CDM model. Note, $\Lambda$CDM tensions \cite{DiValentino:2021izs, Perivolaropoulos:2021jda, Abdalla:2022yfr} independently demand as much. 

The reader may be confused why we have focused on frequentist over Bayesian methods. The motivation can be traced to a mathematical feature of the $\Lambda$CDM model. In exclusively high redshift bins, one encounters pronounced degeneracies when one fits the model to data constraining $H(z)$ or $D_{L}(z) \propto D_{A}(z)$ \cite{Colgain:2022tql}. This leads to banana-shaped contours in 2D MCMC posteriors in the $(H_0, \Omega_m)$-plane and inevitably projection effects. Another problem is that 2D posteriors may only be constrained by the priors leading to prior-dependent results. See Fig. 2 of \cite{Malekjani:2023dky} for an illustration. Furthermore, frequentist methods are computationally cheaper in the setting we consider. Finally, in \cite{Colgain:2023bge} a direct comparison of profile likelihoods and MCMC is made for high redshift $H(z)$ data, where it is noted that frequentist confidence intervals are typically smaller and the profile likelihood allows one to see differences in the $\chi^2$ across the curve in the $(H_0, \Omega_m$)-plane specified by the degeneracy, permitting one to distinguish points in parameter space using frequentist methods that are indistinguishable with Bayesian methods.

Nevertheless, even with profile likelihoods estimating confidence intervals is still tricky. However, in the large sample limit, our results are protected by Wilks' theorem \cite{Wilks}, which is valid when profile likelihoods are close to Gaussian. This applies to the blue curve in Fig. \ref{fig:DES_OM}, where we see a $2.5 \sigma$ discrepancy with Planck. For scientists proficient in statistics, this circumvents a need for MCMC confirmation. However, Fig. \ref{fig:DES_corner} leaves little doubt that one recovers the same trends with MCMC methods, as expected. Ultimately, our work provides an alternative perspective on a $\gtrsim 2 \sigma$ deviation from $\Lambda$CDM reported by DES \cite{DES:2024tys}. A key take-home message here is that deviations from $\Lambda$CDM model are not only evident in extended models, but they must be evident in the baseline model when the data is binned by redshift. See \cite{Colgain:2024mtg} where this statement is put beyond doubt.

Given that we assume the $\Lambda$CDM model, yet find changes in $\Omega_m$ with redshift, it is unclear if the problem concerns the pressureless matter or the DE sector. Nevertheless, we see greater deviations from the Planck $\Omega_m$ value at higher redshifts, where DE is less relevant, so to first approximation this implies a problem in the matter dominated regime. From analysis in \cite{Colgain:2024mtg}, it is clear that one can get an increasing $\Omega_m$ trend from a $w_0 w_a$CDM dynamical DE model at higher redshifts. That being said, there may be pronounced departures from constant $\Omega_m$ at lower redshifts, which need to be validated by both SNe and BAO/full-shape observations. An added problem here is that $w_0 > -1$ \cite{DESI:2024mwx} hinders the ability of dynamical DE to resolve $H_0$ tension \cite{Vagnozzi:2018jhn, Vagnozzi:2019ezj, Alestas:2020mvb, Lee:2022cyh}, which also points us back to the pressureless matter sector. Moreover, the models in \cite{Abdalla:2022yfr} have failed to convincingly resolve the assorted $\Lambda$CDM tensions and a common point in the majority of these models is a pressureless matter sector.

Although pressureless matter can be tested by ensuring that $H(z)$ scales as $H(z) \sim (1+z)^{\frac{3}{2}}$ at high redshifts in the matter-dominated regime, dynamical DE can be parameterised through a function $f(z)$ in the Hubble parameter, $H(z) = H_0 \sqrt{ (1-\Omega_m) f(z) + \Omega_m (1+z)^3}$,  where $f(0)=1$ and at large $z$, $f(z)\sim z^a, \ a<2$. $f(z)$ in principle has an infinite number of degrees of freedom and is hence an unfalsifiable paradigm. For this reason, we must first make sure that matter is pressureless before one buys a function $f(z)$ in $H(z)$. See  \cite{DESI:2024kob, camilleri2024dark} for a sample of functions $f(z)$ confronted to DES SNe that are discrepant with $\Lambda$CDM behaviour. \textit{A priori}, nothing prevents negative DE densities, $f(z) < 0$, which appear in our work as an $\Omega_m > 1$ MLE (see also \cite{Malekjani:2023dky}), and we have witnessed an uptick in recent attempts to model the physics through negative cosmological constants, including cosmological constants that change sign at higher redshifts \cite{Dutta:2018vmq, Visinelli:2019qqu,Calderon:2020hoc,  Acquaviva:2021jov, Akarsu:2022typ, Akarsu:2023mfb, Akarsu:2024qsi,   DESI:2024aqx, wang2024recent}. Such models are supported by an $\Omega_m$ that increases with effective redshift provided one crosses the $\Omega_m = 1$ threshold separating positive and negative DE densities.

\begin{acknowledgments}
This article/publication is based upon work from COST Action CA21136 – “Addressing observational tensions in cosmology with systematics and fundamental physics (CosmoVerse)”, supported by COST (European Cooperation in Science and Technology). SP would like to acknowledge the support of the Iran National Science Foundation (INSF) postdoctoral funds under project No. 4024802. 
MMSHJ  acknowledges support from the ICTP through the senior Associates Programme (2023-2028) and as well as funds from ICTP HECAP section.
\end{acknowledgments}

\appendix 

\section{Non-flat \texorpdfstring{$\Lambda$CDM}{Omegak}}
%\section{Comments on \texorpdfstring{$\Omega_k$}{Omegak}}

In this appendix we explore the implications for curvature $\Omega_k$ of varying the effective redshift of the DES 5YR sample. The motivation for studying curvature is that it scales as $(1+z)^{2}$, versus $(1+z)^{3}$ for pressureless matter and $(1+z)^{0}$ for the cosmological constant, which makes it more relevant than DE at higher redshifts in the late Universe. Explicitly, the non-flat $\Lambda$CDM Hubble parameter is
\begin{equation}
    H(z) = H_0 \sqrt{ 1- \Omega_m -\Omega_k + \Omega_k (1+z)^2 + \Omega_m (1+z)^3},  
\end{equation}
where one should now replace $\chi(z)$ in \eqref{DLz-DMz} with
\begin{equation}\label{chi-k-z}
    \chi_k(z) :=  \left\{ \begin{array}{cc}
    &\frac{1}{\sqrt{\Omega_k}} \sinh \left( \sqrt{\Omega_k} \, \chi(z) \right) \qquad \Omega_k > 0 \\
     &\chi(z) \qquad \Omega_k = 0  \\
     &\frac{1}{\sqrt{-\Omega_k}} \sin \left( \sqrt{-\Omega_k} \, \chi(z) \right) \qquad \Omega_k < 0
    \end{array}\right.
\end{equation}
where $\chi(z)$ is still defined in \eqref{DLz-DMz}. 

We will see that the DES sample strongly embraces a closed Universe at higher redshifts. How this is connected to CMB anomalies pointing to $\Omega_k < 0$ \cite{Handley:2019tkm, DiValentino:2019qzk} is an open question. Nevertheless, here the interpretation is relatively clear. As noted in \cite{Luongo:2021nqh}, the Risaliti-Lusso QSOs \cite{Risaliti:2018reu, Lusso:2020pdb} prefer $\Omega_k < 0$ at higher redshift, e. g. \cite{Khadka:2020tlm}, as a means of decreasing the luminosity distance relative to the Planck flat-$\Lambda$CDM case (due to the sine function in \eqref{chi-k-z}); a similar effect should be at work here.     

\begin{figure}[htb]
   \centering
\includegraphics[width=90mm]{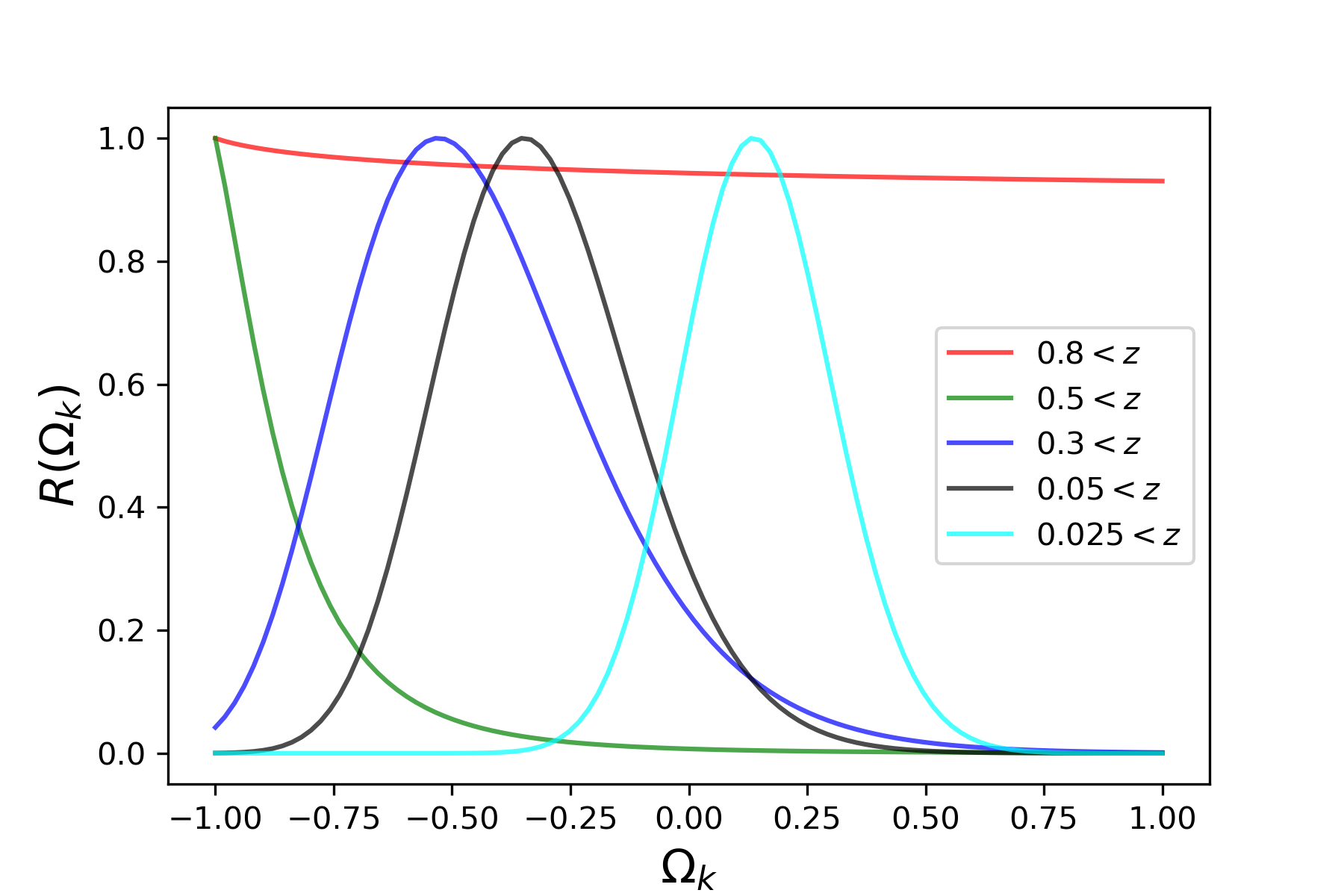} 
\caption{Profile likelihood ratios for curvature $\Omega_k$ with increasing effective redshift. The cyan and black profiles differ only in the addition of SNe from external data sets to the DES 5YR SNe sample.}
\label{fig:DES_OK}
\end{figure}

In Fig. \ref{fig:DES_OK} we show the profile likelihood ratios, where cyan denotes the full DES 5YR sample including the 194 CfA/CSP/Foundation SNe \cite{Hicken:2009df, Hicken:2012zr, Krisciunas:2017yoe, Foley:2017zdq}, whereas black corresponds to the 1635 DES SNe on their own. To construct the plot we have imposed  uniform priors $H_0 \in [0, 200]$, $\Omega_m \in [0, 1]$, while scanning over $\Omega_k$ in the range $\Omega_k \in [-1, 1]$. Note, our priors are wider than the DES priors $\Omega_k \in [-0.5, 0.5]$ \cite{DES:2024tys}, and we have made sure when we minimise the $\chi^2$ with respect to $(H_0, \Omega_m)$ that the MLE are not impacted by our bounds on the auxiliary parameters $(H_0, \Omega_m)$. {We remark also that since our prior is not physically motivated and there is no boundary between our MLEs and $\Omega_k=0$, the Feldman-Cousins prescription \cite{Feldman:1997qc} is not relevant.} 

The first thing to note from Table \ref{tab:OKvsZ} is that when we use the full sample, we agree with Table 2 of \cite{DES:2024tys} that $\Omega_m = 0.16 \pm 0.16$. Slight disagreements are due to the fact that we have only sampled $\Omega_k$ at approximately 100 evenly spaced $\Omega_k$ values and we expect that one can improve agreement by increasing the number of $\Omega_k$ where we minimise the $\chi^2$. Thus, the cyan profile likelihood in Fig. \ref{fig:DES_OK} is in line with expectations. Nevertheless, as we remove the low redshift SNe, starting with the 194 SNe common to the CfA/CSP/Foundation samples \cite{Hicken:2009df, Hicken:2012zr, Krisciunas:2017yoe, Foley:2017zdq}, the profile likelihood peak or MLE shifts to negative $\Omega_k$ values. Noting that $\Omega_k = 0$ recovers our findings in the main text, we recognise that the fit to data can be improved through $\Omega_k < 0$. For the highest redshift SNe $z > 0.8$, the $\chi^2$ improvement is not great, as is evident from the red curve in Fig. \ref{fig:DES_OK}. Nevertheless, the improvement is non-negligible at lower redshifts and is extremely pronounced for DES SNe with $z > 0.5$, as demonstrated by the green curve in Fig. \ref{fig:DES_OK}. From Table \ref{tab:OKvsZ}, once one excludes the lowest redshift SNe in the CfA/CSP/Foundation samples, it is evident that DES SNe on their own prefer a closed Universe with $\Omega_k < 0$.

We can drill down on this discrepancy a bit more. Invoking Wilks' theorem \cite{Wilks} and following a similar analysis as in the main text, a flat Universe appears at $\Delta \chi^2 = 0.89, 2.5, 3.1, 10.0, 0.1$, corresponding to $0.9 \sigma, 1.6 \sigma, 1.8 \sigma, 3.2 \sigma, 0.3 \sigma$ for the cyan, black, blue, green and red profile likelihoods, respectively. For the lowest redshifts probed in Fig. \ref{fig:DES_OK} it is clear that profile likelihoods are close to Gaussian, so we expect this to be a good approximation. For $z > 0.5$ our boundary at $\Omega_k = -1$ prevents us from finding the MLE, but it is obvious that it appears at $\Omega_k < -1$. Nevertheless, assuming that the profile likelihood peak appears at $\Omega_k = -1$ allows us to get a \textit{lower bound} on any discrepancy with the flat Universe. This bounds the preference for a closed Universe from DES SNe with $z > 0.5$ below at $3.2 \sigma$. Note, there is nothing unphysical about $\Omega_k < -1$ parameter space, so one could simply relax the priors, as we have done in the text, to find the MLE or profile likelihood peak, which would invariably exacerbate the tension.

\begin{table}[htb]
\centering 
\begin{tabular}{c|c|c|c}
 \rule{0pt}{3ex} Redshift & \# SNe & $ \Omega_{k}$ ($68 \%$) & $\Omega_k$ ($\Delta \chi^2 \leq 1$) \\
 \hline 
\rule{0pt}{3ex} $0.025 < z$ & $1829$ & $0.13^{+0.16}_{-0.16}$ & $0.13^{+0.16}_{-0.14}$ \\
\hline 
\rule{0pt}{3ex} $0.05 < z$ & $1635$ & $-0.35^{+0.24}_{-0.18}$ & $-0.35^{+0.22}_{-0.18}$ \\
\hline
\rule{0pt}{3ex} $0.3 < z$  & $1426$ & $-0.54^{+0.30}_{-0.22}$ & $-0.54^{+0.28}_{-0.22}$ \\
\hline
\rule{0pt}{3ex} $0.5 < z$ & $874$ & $< - 0.80$ ($-1$) & $ < -0.91$ ($-1$) \\
\hline
\rule{0pt}{3ex} $ 0.8 < z$ & $128$ & $<1$ ($-1$) & $ <1$ ($-1$)  \\
\end{tabular}
\caption{$68 \%$ confidence intervals inferred with two methods from DES SNe in different redshift ranges. The number of SNe are also documented. For one-sided confidence intervals we record the MLE in brackets.}
\label{tab:OKvsZ}
\end{table}

We can get an alternative handle on the discrepancies by integrating under the (normalised) profile likelihood ratio curve as outlined in the text. Doing so, we find a flat Universe is located at $ 0.9\sigma, 1.6 \sigma, 1.7 \sigma, 2.5 \sigma, 0.7 \sigma$ for the cyan, black, blue, green and red profile likelihoods, respectively. Where we find good agreement between the methods, it is expected since Wilks' theorem assumes profile likelihoods are Gaussian, an assumption that is valid in the large sample limit. Evidently,  $1426$, $1635$ and $1829$ SNe samples  (see Table \ref{tab:OKvsZ}) are large enough, since the profile likelihood are visibly Gaussian in Fig. \ref{fig:DES_OK}. Samples of $128$ and $874$ SNe may also be large samples and whether Wilks' theorem is a valid approximation  can only be judged by relaxing the $\Omega_k$ priors to locate the MLE or profile likelihood peak. Understandably, the greatest difference in statistical significance between the profile likelihood ratio confidence levels arises where the profile likelihood is impacted by the $\Omega_k =-1$ bound. However, what is important here is that different methods can distinguish insignificant discrepancies $( \lesssim 1 \sigma)$ from more significant tensions $(\gtrsim 2.5 \sigma)$.

\bibliographystyle{fullsort.bst}
\bibliography{refs}

\providecommand{\href}[2]{#2}\begingroup\raggedright\begin{thebibliography}{10}

\bibitem{DiValentino:2021izs}
E.~Di~Valentino, O.~Mena, S.~Pan, L.~Visinelli, W.~Yang, A.~Melchiorri, D.~F. Mota, A.~G. Riess, and J.~Silk, ``{In the realm of the Hubble tension\textemdash{}a review of solutions},'' {\em Class. Quant. Grav.} {\bf 38} (2021), no.~15, 153001, \href{http://www.arXiv.org/abs/2103.01183}{{\tt 2103.01183}}.

\bibitem{Perivolaropoulos:2021jda}
L.~Perivolaropoulos and F.~Skara, ``{Challenges for \ensuremath{\Lambda}CDM: An update},'' {\em New Astron. Rev.} {\bf 95} (2022) 101659, \href{http://www.arXiv.org/abs/2105.05208}{{\tt 2105.05208}}.

\bibitem{Abdalla:2022yfr}
E.~Abdalla {\em et al.}, ``{Cosmology intertwined: A review of the particle physics, astrophysics, and cosmology associated with the cosmological tensions and anomalies},'' {\em JHEAp} {\bf 34} (2022) 49--211, \href{http://www.arXiv.org/abs/2203.06142}{{\tt 2203.06142}}.

\bibitem{Krishnan:2020vaf}
C.~Krishnan, E.~\'O~Colg\'ain, M.~M. Sheikh-Jabbari, and T.~Yang, ``{Running Hubble Tension and a H0 Diagnostic},'' {\em Phys. Rev. D} {\bf 103} (2021), no.~10, 103509, \href{http://www.arXiv.org/abs/2011.02858}{{\tt 2011.02858}}.

\bibitem{Krishnan:2022fzz}
C.~Krishnan and R.~Mondol, ``{$H_0$ as a Universal FLRW Diagnostic},'' \href{http://www.arXiv.org/abs/2201.13384}{{\tt 2201.13384}}.

\bibitem{Wong:2019kwg}
K.~C. Wong {\em et al.}, ``{H0LiCOW \textendash{} XIII. A 2.4 per cent measurement of H0 from lensed quasars: 5.3\ensuremath{\sigma} tension between early- and late-Universe probes},'' {\em Mon. Not. Roy. Astron. Soc.} {\bf 498} (2020), no.~1, 1420--1439, \href{http://www.arXiv.org/abs/1907.04869}{{\tt 1907.04869}}.

\bibitem{Millon:2019slk}
M.~Millon {\em et al.}, ``{TDCOSMO. I. An exploration of systematic uncertainties in the inference of $H_0$ from time-delay cosmography},'' {\em Astron. Astrophys.} {\bf 639} (2020) A101, \href{http://www.arXiv.org/abs/1912.08027}{{\tt 1912.08027}}.

\bibitem{Krishnan:2020obg}
C.~Krishnan, E.~\'O~Colg\'ain, Ruchika, A.~A. Sen, M.~M. Sheikh-Jabbari, and T.~Yang, ``{Is there an early Universe solution to Hubble tension?},'' {\em Phys. Rev. D} {\bf 102} (2020), no.~10, 103525, \href{http://www.arXiv.org/abs/2002.06044}{{\tt 2002.06044}}.

\bibitem{Dainotti:2021pqg}
M.~G. Dainotti, B.~De~Simone, T.~Schiavone, G.~Montani, E.~Rinaldi, and G.~Lambiase, ``{On the Hubble constant tension in the SNe Ia Pantheon sample},'' {\em Astrophys. J.} {\bf 912} (2021), no.~2, 150, \href{http://www.arXiv.org/abs/2103.02117}{{\tt 2103.02117}}.

\bibitem{Dainotti:2022bzg}
M.~G. Dainotti, B.~De~Simone, T.~Schiavone, G.~Montani, E.~Rinaldi, G.~Lambiase, M.~Bogdan, and S.~Ugale, ``{On the Evolution of the Hubble Constant with the SNe Ia Pantheon Sample and Baryon Acoustic Oscillations: A Feasibility Study for GRB-Cosmology in 2030},'' {\em Galaxies} {\bf 10} (2022), no.~1, 24, \href{http://www.arXiv.org/abs/2201.09848}{{\tt 2201.09848}}.

\bibitem{Hu:2022kes}
J.-P. Hu and F.~Y. Wang, ``{Revealing the late-time transition of H0: relieve the Hubble crisis},'' {\em Mon. Not. Roy. Astron. Soc.} {\bf 517} (2022), no.~1, 576--581, \href{http://www.arXiv.org/abs/2203.13037}{{\tt 2203.13037}}.

\bibitem{Colgain:2022nlb}
E.~\'O~Colg\'ain, M.~M. Sheikh-Jabbari, R.~Solomon, G.~Bargiacchi, S.~Capozziello, M.~G. Dainotti, and D.~Stojkovic, ``{Revealing intrinsic flat \ensuremath{\Lambda}CDM biases with standardizable candles},'' {\em Phys. Rev. D} {\bf 106} (2022), no.~4, L041301, \href{http://www.arXiv.org/abs/2203.10558}{{\tt 2203.10558}}.

\bibitem{Colgain:2022rxy}
E.~\'O~Colg\'ain, M.~M. Sheikh-Jabbari, R.~Solomon, M.~G. Dainotti, and D.~Stojkovic, ``{Putting flat \ensuremath{\Lambda}CDM in the (Redshift) bin},'' {\em Phys. Dark Univ.} {\bf 44} (2024) 101464, \href{http://www.arXiv.org/abs/2206.11447}{{\tt 2206.11447}}.

\bibitem{Jia:2022ycc}
X.~D. Jia, J.~P. Hu, and F.~Y. Wang, ``{Evidence of a decreasing trend for the Hubble constant},'' {\em Astron. Astrophys.} {\bf 674} (2023) A45, \href{http://www.arXiv.org/abs/2212.00238}{{\tt 2212.00238}}.

\bibitem{jia2024uncorrelated}
X.~D. Jia, J.~P. Hu, and F.~Y. Wang, ``{Uncorrelated estimations of $H_0$ redshift evolution from DESI baryon acoustic oscillation observations},'' \href{http://www.arXiv.org/abs/2406.02019}{{\tt 2406.02019}}.

\bibitem{Risaliti:2018reu}
G.~Risaliti and E.~Lusso, ``{Cosmological constraints from the Hubble diagram of quasars at high redshifts},'' {\em Nature Astron.} {\bf 3} (2019), no.~3, 272--277, \href{http://www.arXiv.org/abs/1811.02590}{{\tt 1811.02590}}.

\bibitem{Lusso:2020pdb}
E.~Lusso {\em et al.}, ``{Quasars as standard candles III. Validation of a new sample for cosmological studies},'' {\em Astron. Astrophys.} {\bf 642} (2020) A150, \href{http://www.arXiv.org/abs/2008.08586}{{\tt 2008.08586}}.

\bibitem{Pourojaghi:2022zrh}
S.~Pourojaghi, N.~F. Zabihi, and M.~Malekjani, ``{Can high-redshift Hubble diagrams rule out the standard model of cosmology in the context of cosmography?},'' {\em Phys. Rev. D} {\bf 106} (2022), no.~12, 123523, \href{http://www.arXiv.org/abs/2212.04118}{{\tt 2212.04118}}.

\bibitem{Pasten:2023rpc}
E.~Past\'en and V.~H. C\'ardenas, ``{Testing \ensuremath{\Lambda}CDM cosmology in a binned universe: Anomalies in the deceleration parameter},'' {\em Phys. Dark Univ.} {\bf 40} (2023) 101224, \href{http://www.arXiv.org/abs/2301.10740}{{\tt 2301.10740}}.

\bibitem{Akarsu:2024qiq}
O.~Akarsu, E.~\'O~Colg\'ain, A.~A. Sen, and M.~M. Sheikh-Jabbari, ``{$\Lambda$CDM Tensions: Localising Missing Physics through Consistency Checks},'' \href{http://www.arXiv.org/abs/2402.04767}{{\tt 2402.04767}}.

\bibitem{Risaliti:2015zla}
G.~Risaliti and E.~Lusso, ``{A Hubble Diagram for Quasars},'' {\em Astrophys. J.} {\bf 815} (2015) 33, \href{http://www.arXiv.org/abs/1505.07118}{{\tt 1505.07118}}.

\bibitem{colgain2024high}
E.~{{\'O} Colg{\'a}in}, M.~M. {Sheikh-Jabbari}, and L.~{Yin}, ``{Do high redshift QSOs and GRBs corroborate JWST?},'' {\em arXiv e-prints} (May, 2024) arXiv:2405.19953, \href{http://www.arXiv.org/abs/2405.19953}{{\tt 2405.19953}}.

\bibitem{Yang:2019vgk}
T.~Yang, A.~Banerjee, and E.~\'O~Colg\'ain, ``{Cosmography and flat $\Lambda$CDM tensions at high redshift},'' {\em Phys. Rev. D} {\bf 102} (2020), no.~12, 123532, \href{http://www.arXiv.org/abs/1911.01681}{{\tt 1911.01681}}.

\bibitem{Khadka:2020vlh}
N.~Khadka and B.~Ratra, ``{Using quasar X-ray and UV flux measurements to constrain cosmological model parameters},'' {\em Mon. Not. Roy. Astron. Soc.} {\bf 497} (2020), no.~1, 263--278, \href{http://www.arXiv.org/abs/2004.09979}{{\tt 2004.09979}}.

\bibitem{Khadka:2020tlm}
N.~Khadka and B.~Ratra, ``{Determining the range of validity of quasar X-ray and UV flux measurements for constraining cosmological model parameters},'' {\em Mon. Not. Roy. Astron. Soc.} {\bf 502} (2021), no.~4, 6140--6156, \href{http://www.arXiv.org/abs/2012.09291}{{\tt 2012.09291}}.

\bibitem{Riess:2021jrx}
A.~G. Riess {\em et al.}, ``{A Comprehensive Measurement of the Local Value of the Hubble Constant with 1 km/s/Mpc Uncertainty from the Hubble Space Telescope and the SH0ES Team},'' {\em Astrophys. J. Lett.} {\bf 934} (2022), no.~1, L7, \href{http://www.arXiv.org/abs/2112.04510}{{\tt 2112.04510}}.

\bibitem{Khadka:2021xcc}
N.~Khadka and B.~Ratra, ``{Do quasar X-ray and UV flux measurements provide a useful test of cosmological models?},'' {\em Mon. Not. Roy. Astron. Soc.} {\bf 510} (2022), no.~2, 2753--2772, \href{http://www.arXiv.org/abs/2107.07600}{{\tt 2107.07600}}.

\bibitem{Khadka:2022aeg}
N.~Khadka, M.~Zaja\v{c}ek, R.~Prince, S.~Panda, B.~Czerny, M.~L. Mart\'\i{}nez-Aldama, V.~K. Jaiswal, and B.~Ratra, ``{Quasar UV/X-ray relation luminosity distances are shorter than reverberation-measured radius\textendash{}luminosity relation luminosity distances},'' {\em Mon. Not. Roy. Astron. Soc.} {\bf 522} (2023), no.~1, 1247--1264, \href{http://www.arXiv.org/abs/2212.10483}{{\tt 2212.10483}}.

\bibitem{Singal:2022nto}
J.~Singal, S.~Mutchnick, and V.~Petrosian, ``{The X-Ray Luminosity Function Evolution of Quasars and the Correlation between the X-Ray and Ultraviolet Luminosities},'' {\em Astrophys. J.} {\bf 932} (2022), no.~2, 111, \href{http://www.arXiv.org/abs/2203.13374}{{\tt 2203.13374}}.

\bibitem{Petrosian:2022tlp}
V.~Petrosian, J.~Singal, and S.~Mutchnick, ``{Can the Distance-Redshift Relation be Determined from Correlations between Luminosities?},'' {\em Astrophys. J. Lett.} {\bf 935} (2022), no.~1, L19, \href{http://www.arXiv.org/abs/2205.07981}{{\tt 2205.07981}}.

\bibitem{Zajacek:2023qjm}
M.~Zaja\v{c}ek, B.~Czerny, N.~Khadka, M.~L. Mart\'\i{}nez-Aldama, R.~Prince, S.~Panda, and B.~Ratra, ``{Effect of Extinction on Quasar Luminosity Distances Determined from UV and X-Ray Flux Measurements},'' {\em Astrophys. J.} {\bf 961} (2024), no.~2, 229, \href{http://www.arXiv.org/abs/2305.08179}{{\tt 2305.08179}}.

\bibitem{Malekjani:2023dky}
M.~Malekjani, R.~M. Conville, E.~\'O~Colg\'ain, S.~Pourojaghi, and M.~M. Sheikh-Jabbari, ``{On redshift evolution and negative dark energy density in Pantheon + Supernovae},'' {\em Eur. Phys. J. C} {\bf 84} (2024), no.~3, 317, \href{http://www.arXiv.org/abs/2301.12725}{{\tt 2301.12725}}.

\bibitem{Pan-STARRS1:2017jku}
{\bf Pan-STARRS1} Collaboration, D.~M. Scolnic {\em et al.}, ``{The Complete Light-curve Sample of Spectroscopically Confirmed SNe Ia from Pan-STARRS1 and Cosmological Constraints from the Combined Pantheon Sample},'' {\em Astrophys. J.} {\bf 859} (2018), no.~2, 101, \href{http://www.arXiv.org/abs/1710.00845}{{\tt 1710.00845}}.

\bibitem{Scolnic:2021amr}
D.~Scolnic {\em et al.}, ``{The Pantheon+ Analysis: The Full Data Set and Light-curve Release},'' {\em Astrophys. J.} {\bf 938} (2022), no.~2, 113, \href{http://www.arXiv.org/abs/2112.03863}{{\tt 2112.03863}}.

\bibitem{Brout:2022vxf}
D.~Brout {\em et al.}, ``{The Pantheon+ Analysis: Cosmological Constraints},'' {\em Astrophys. J.} {\bf 938} (2022), no.~2, 110, \href{http://www.arXiv.org/abs/2202.04077}{{\tt 2202.04077}}.

\bibitem{SNLS:2011lii}
{\bf SNLS} Collaboration, A.~Conley {\em et al.}, ``{Supernova Constraints and Systematic Uncertainties from the First 3 Years of the Supernova Legacy Survey},'' {\em Astrophys. J. Suppl.} {\bf 192} (2011) 1, \href{http://www.arXiv.org/abs/1104.1443}{{\tt 1104.1443}}.

\bibitem{SNLS:2011cra}
{\bf SNLS} Collaboration, M.~Sullivan {\em et al.}, ``{SNLS3: Constraints on Dark Energy Combining the Supernova Legacy Survey Three Year Data with Other Probes},'' {\em Astrophys. J.} {\bf 737} (2011) 102, \href{http://www.arXiv.org/abs/1104.1444}{{\tt 1104.1444}}.

\bibitem{Brout:2021mpj}
D.~Brout {\em et al.}, ``{The Pantheon+ Analysis: SuperCal-fragilistic Cross Calibration, Retrained SALT2 Light-curve Model, and Calibration Systematic Uncertainty},'' {\em Astrophys. J.} {\bf 938} (2022), no.~2, 111, \href{http://www.arXiv.org/abs/2112.03864}{{\tt 2112.03864}}.

\bibitem{DES:2024tys}
{\bf DES} Collaboration, T.~M.~C. Abbott and et~al, ``The dark energy survey: Cosmology results with ~1500 new high-redshift type ia supernovae using the full 5-year dataset,'' \href{http://www.arXiv.org/abs/2401.02929}{{\tt 2401.02929}}.

\bibitem{DES:2024upw}
{\bf DES} Collaboration, B.~O. S\'anchez {\em et al.}, ``{The Dark Energy Survey Supernova Program: Light curves and 5-Year data release},'' \href{http://www.arXiv.org/abs/2406.05046}{{\tt 2406.05046}}.

\bibitem{the_des_sn_working_group_2024_12720778}
T.~D. S.~W. Group, ``Des supernova 5yr data release,'' July, 2024.

\bibitem{Rubin:2023ovl}
D.~Rubin {\em et al.}, ``{Union Through UNITY: Cosmology with 2,000 SNe Using a Unified Bayesian Framework},'' \href{http://www.arXiv.org/abs/2311.12098}{{\tt 2311.12098}}.

\bibitem{DESI:2024mwx}
{\bf DESI} Collaboration, A.~G. Adame {\em et al.}, ``{DESI 2024 VI: Cosmological Constraints from the Measurements of Baryon Acoustic Oscillations},'' \href{http://www.arXiv.org/abs/2404.03002}{{\tt 2404.03002}}.

\bibitem{Planck:2018vyg}
{\bf Planck} Collaboration, N.~Aghanim {\em et al.}, ``{Planck 2018 results. VI. Cosmological parameters},'' {\em Astron. Astrophys.} {\bf 641} (2020) A6, \href{http://www.arXiv.org/abs/1807.06209}{{\tt 1807.06209}}. [Erratum: Astron.Astrophys. 652, C4 (2021)].

\bibitem{Trotta:2017wnx}
R.~Trotta, ``{Bayesian Methods in Cosmology},'' \href{http://www.arXiv.org/abs/1701.01467}{{\tt 1701.01467}}.

\bibitem{Wilks}
S.~S. Wilks, ``{The Large-Sample Distribution of the Likelihood Ratio for Testing Composite Hypotheses},'' {\em The Annals of Mathematical Statistics} {\bf 9} (1938), no.~1, 60 -- 62.

\bibitem{Gomez-Valent:2022hkb}
A.~G\'omez-Valent, ``{Fast test to assess the impact of marginalization in Monte~Carlo analyses and its application to cosmology},'' {\em Phys. Rev. D} {\bf 106} (2022), no.~6, 063506, \href{http://www.arXiv.org/abs/2203.16285}{{\tt 2203.16285}}.

\bibitem{Colgain:2023bge}
E.~\'O~Colg\'ain, S.~Pourojaghi, M.~M. Sheikh-Jabbari, and D.~Sherwin, ``{MCMC Marginalisation Bias and $\Lambda$CDM tensions},'' \href{http://www.arXiv.org/abs/2307.16349}{{\tt 2307.16349}}.

\bibitem{Hicken:2009df}
M.~Hicken, P.~Challis, S.~Jha, R.~P. Kirsher, T.~Matheson, M.~Modjaz, A.~Rest, and W.~M. Wood-Vasey, ``{CfA3: 185 Type Ia Supernova Light Curves from the CfA},'' {\em Astrophys. J.} {\bf 700} (2009) 331--357, \href{http://www.arXiv.org/abs/0901.4787}{{\tt 0901.4787}}.

\bibitem{Hicken:2012zr}
M.~Hicken {\em et al.}, ``{CfA4: Light Curves for 94 Type Ia Supernovae},'' {\em Astrophys. J. Suppl.} {\bf 200} (2012) 12, \href{http://www.arXiv.org/abs/1205.4493}{{\tt 1205.4493}}.

\bibitem{Krisciunas:2017yoe}
K.~Krisciunas {\em et al.}, ``{The Carnegie Supernova Project I: Third Photometry Data Release of Low-Redshift Type Ia Supernovae and Other White Dwarf Explosions},'' {\em Astron. J.} {\bf 154} (2017), no.~5, 211, \href{http://www.arXiv.org/abs/1709.05146}{{\tt 1709.05146}}.

\bibitem{Foley:2017zdq}
R.~J. Foley {\em et al.}, ``{The Foundation Supernova Survey: Motivation, Design, Implementation, and First Data Release},'' {\em Mon. Not. Roy. Astron. Soc.} {\bf 475} (2018), no.~1, 193--219, \href{http://www.arXiv.org/abs/1711.02474}{{\tt 1711.02474}}.

\bibitem{Herold:2021ksg}
L.~Herold, E.~G.~M. Ferreira, and E.~Komatsu, ``{New Constraint on Early Dark Energy from Planck and BOSS Data Using the Profile Likelihood},'' {\em Astrophys. J. Lett.} {\bf 929} (2022), no.~1, L16, \href{http://www.arXiv.org/abs/2112.12140}{{\tt 2112.12140}}.

\bibitem{Feldman:1997qc}
G.~J. Feldman and R.~D. Cousins, ``{A Unified approach to the classical statistical analysis of small signals},'' {\em Phys. Rev. D} {\bf 57} (1998) 3873--3889, \href{http://www.arXiv.org/abs/physics/9711021}{{\tt physics/9711021}}.

\bibitem{Weinberg:1988cp}
S.~Weinberg, ``{The Cosmological Constant Problem},'' {\em Rev. Mod. Phys.} {\bf 61} (1989) 1--23.

\bibitem{Dvali:2014gua}
G.~Dvali and C.~Gomez, ``{Quantum Exclusion of Positive Cosmological Constant?},'' {\em Annalen Phys.} {\bf 528} (2016) 68--73, \href{http://www.arXiv.org/abs/1412.8077}{{\tt 1412.8077}}.

\bibitem{Dvali:2018fqu}
G.~Dvali and C.~Gomez, ``{On Exclusion of Positive Cosmological Constant},'' {\em Fortsch. Phys.} {\bf 67} (2019), no.~1-2, 1800092, \href{http://www.arXiv.org/abs/1806.10877}{{\tt 1806.10877}}.

\bibitem{Obied:2018sgi}
G.~{Obied}, H.~{Ooguri}, L.~{Spodyneiko}, and C.~{Vafa}, ``{De Sitter Space and the Swampland},'' {\em arXiv e-prints} (June, 2018) arXiv:1806.08362, \href{http://www.arXiv.org/abs/1806.08362}{{\tt 1806.08362}}.

\bibitem{Colgain:2024mtg}
E.~{\'O Colg{\'a}in} and M.~M. {Sheikh-Jabbari}, ``{DESI and SNe: Dynamical Dark Energy, $\Omega_m$ Tension or Systematics?},'' {\em arXiv e-prints} (Dec., 2024) arXiv:2412.12905, \href{http://www.arXiv.org/abs/2412.12905}{{\tt 2412.12905}}.

\bibitem{Moresco:2022phi}
M.~Moresco {\em et al.}, ``{Unveiling the Universe with emerging cosmological probes},'' {\em Living Rev. Rel.} {\bf 25} (2022), no.~1, 6, \href{http://www.arXiv.org/abs/2201.07241}{{\tt 2201.07241}}.

\bibitem{Efstathiou:2024xcq}
G.~Efstathiou, ``{Evolving Dark Energy or Supernovae Systematics?},'' \href{http://www.arXiv.org/abs/2408.07175}{{\tt 2408.07175}}.

\bibitem{Colgain:2022tql}
E.~\'O~Colg\'ain, M.~M. Sheikh-Jabbari, and R.~Solomon, ``{High redshift \ensuremath{\Lambda}CDM cosmology: To bin or not to bin?},'' {\em Phys. Dark Univ.} {\bf 40} (2023) 101216, \href{http://www.arXiv.org/abs/2211.02129}{{\tt 2211.02129}}.

\bibitem{Vagnozzi:2018jhn}
S.~Vagnozzi, S.~Dhawan, M.~Gerbino, K.~Freese, A.~Goobar, and O.~Mena, ``{Constraints on the sum of the neutrino masses in dynamical dark energy models with $w(z) \geq -1$ are tighter than those obtained in $\Lambda$CDM},'' {\em Phys. Rev. D} {\bf 98} (2018), no.~8, 083501, \href{http://www.arXiv.org/abs/1801.08553}{{\tt 1801.08553}}.

\bibitem{Vagnozzi:2019ezj}
S.~Vagnozzi, ``{New physics in light of the $H_0$ tension: An alternative view},'' {\em Phys. Rev. D} {\bf 102} (2020), no.~2, 023518, \href{http://www.arXiv.org/abs/1907.07569}{{\tt 1907.07569}}.

\bibitem{Alestas:2020mvb}
G.~Alestas, L.~Kazantzidis, and L.~Perivolaropoulos, ``{$H_0$ tension, phantom dark energy, and cosmological parameter degeneracies},'' {\em Phys. Rev. D} {\bf 101} (2020), no.~12, 123516, \href{http://www.arXiv.org/abs/2004.08363}{{\tt 2004.08363}}.

\bibitem{Lee:2022cyh}
B.-H. Lee, W.~Lee, E.~O. Colg\'ain, M.~M. Sheikh-Jabbari, and S.~Thakur, ``{Is local H $_{0}$ at odds with dark energy EFT?},'' {\em JCAP} {\bf 04} (2022), no.~04, 004, \href{http://www.arXiv.org/abs/2202.03906}{{\tt 2202.03906}}.

\bibitem{DESI:2024kob}
{\bf DESI} Collaboration, K.~Lodha {\em et al.}, ``{DESI 2024: Constraints on Physics-Focused Aspects of Dark Energy using DESI DR1 BAO Data},'' \href{http://www.arXiv.org/abs/2405.13588}{{\tt 2405.13588}}.

\bibitem{camilleri2024dark}
{\bf DES} Collaboration, R.~Camilleri {\em et al.}, ``The dark energy survey supernova program: Investigating beyond-$\lambda$cdm,'' \href{http://www.arXiv.org/abs/2406.05048}{{\tt 2406.05048}}.

\bibitem{Dutta:2018vmq}
K.~Dutta, Ruchika, A.~Roy, A.~A. Sen, and M.~M. Sheikh-Jabbari, ``{Beyond $\Lambda $CDM with low and high redshift data: implications for dark energy},'' {\em Gen. Rel. Grav.} {\bf 52} (2020), no.~2, 15, \href{http://www.arXiv.org/abs/1808.06623}{{\tt 1808.06623}}.

\bibitem{Visinelli:2019qqu}
L.~Visinelli, S.~Vagnozzi, and U.~Danielsson, ``{Revisiting a negative cosmological constant from low-redshift data},'' {\em Symmetry} {\bf 11} (2019), no.~8, 1035, \href{http://www.arXiv.org/abs/1907.07953}{{\tt 1907.07953}}.

\bibitem{Calderon:2020hoc}
R.~Calder\'on, R.~Gannouji, B.~L'Huillier, and D.~Polarski, ``{Negative cosmological constant in the dark sector?},'' {\em Phys. Rev. D} {\bf 103} (2021), no.~2, 023526, \href{http://www.arXiv.org/abs/2008.10237}{{\tt 2008.10237}}.

\bibitem{Acquaviva:2021jov}
G.~Acquaviva, O.~Akarsu, N.~Katirci, and J.~A. Vazquez, ``{Simple-graduated dark energy and spatial curvature},'' {\em Phys. Rev. D} {\bf 104} (2021), no.~2, 023505, \href{http://www.arXiv.org/abs/2104.02623}{{\tt 2104.02623}}.

\bibitem{Akarsu:2022typ}
O.~Akarsu, S.~Kumar, E.~\"Oz\"ulker, J.~A. Vazquez, and A.~Yadav, ``{Relaxing cosmological tensions with a sign switching cosmological constant: Improved results with Planck, BAO, and Pantheon data},'' {\em Phys. Rev. D} {\bf 108} (2023), no.~2, 023513, \href{http://www.arXiv.org/abs/2211.05742}{{\tt 2211.05742}}.

\bibitem{Akarsu:2023mfb}
O.~Akarsu, E.~Di~Valentino, S.~Kumar, R.~C. Nunes, J.~A. Vazquez, and A.~Yadav, ``{$\Lambda_{\rm s}$CDM model: A promising scenario for alleviation of cosmological tensions},'' \href{http://www.arXiv.org/abs/2307.10899}{{\tt 2307.10899}}.

\bibitem{Akarsu:2024qsi}
O.~Akarsu, A.~De~Felice, E.~Di~Valentino, S.~Kumar, R.~C. Nunes, E.~Ozulker, J.~A. Vazquez, and A.~Yadav, ``{$\Lambda_{\rm s}$CDM cosmology from a type-II minimally modified gravity},'' \href{http://www.arXiv.org/abs/2402.07716}{{\tt 2402.07716}}.

\bibitem{DESI:2024aqx}
{\bf DESI} Collaboration, R.~Calderon {\em et al.}, ``{DESI 2024: Reconstructing Dark Energy using Crossing Statistics with DESI DR1 BAO data},'' \href{http://www.arXiv.org/abs/2405.04216}{{\tt 2405.04216}}.

\bibitem{wang2024recent}
H.~Wang, Z.-Y. Peng, and Y.-S. Piao, ``Can recent desi bao measurements accommodate a negative cosmological constant?,'' 2024.

\bibitem{Handley:2019tkm}
W.~Handley, ``{Curvature tension: evidence for a closed universe},'' {\em Phys. Rev. D} {\bf 103} (2021), no.~4, L041301, \href{http://www.arXiv.org/abs/1908.09139}{{\tt 1908.09139}}.

\bibitem{DiValentino:2019qzk}
E.~Di~Valentino, A.~Melchiorri, and J.~Silk, ``{Planck evidence for a closed Universe and a possible crisis for cosmology},'' {\em Nature Astron.} {\bf 4} (2019), no.~2, 196--203, \href{http://www.arXiv.org/abs/1911.02087}{{\tt 1911.02087}}.

\bibitem{Luongo:2021nqh}
O.~Luongo, M.~Muccino, E.~\'O~Colg\'ain, M.~M. Sheikh-Jabbari, and L.~Yin, ``{Larger H0 values in the CMB dipole direction},'' {\em Phys. Rev. D} {\bf 105} (2022), no.~10, 103510, \href{http://www.arXiv.org/abs/2108.13228}{{\tt 2108.13228}}.

\end{thebibliography}\endgroup

\end{document}